# Band nonlinearity-enabled manipulation of Dirac nodes, Weyl cones, and valleytronics with intense linearly polarized light

Ofer Neufeld[1,*], Hannes Hübener[1], Gregor Jotzu[1], Umberto De Giovannini[1,2], Angel Rubio[1,3,*]

[1]Max Planck Institute for the Structure and Dynamics of Matter and Center for Free-electron Laser Science, Hamburg 22761, Germany.

[2]Università degli Studi di Palermo, Dipartimento di Fisica e Chimica—Emilio Segrè, Palermo I-90123, Italy.

[3]Center for Computational Quantum Physics (CCQ), The Flatiron Institute, New York 10010, USA.

*Corresponding author E-mails: oneufeld@schmidtsciencefellows.org, angel.rubio@mpsd.mpg.de.

ABSTRACT: We study low-frequency linearly-polarized laser-dressing in materials with valley (graphene and hexagonal-Boron-Nitride), and topological (Dirac- and Weyl-semimetals), properties. In Dirac-like linearly-dispersing bands, the laser substantially moves the Dirac nodes away from their original position, and the movement direction can be fully controlled by rotating the laser polarization. We prove that this effect originates from band nonlinearities away from the Dirac nodes. We further demonstrate that this physical mechanism is widely applicable, and can move the positions of the valley minima in hexagonal materials to tune valley selectivity, split and move Weyl cones in higher-order Weyl semimetals, and merge Dirac nodes in three-dimensional Dirac semimetals. The model results are validated with *ab-initio* calculations. Our results directly affect efforts for exploring light-dressed electronic-structure, suggesting that one can benefit from band nonlinearity for tailoring material properties, and highlight the importance of the full band structure in nonlinear optical phenomena in solids.

Light-induced band structure and opto-electronic device engineering has gained considerable attention in recent years due to its potential to revolutionize electronics [1–21]. Within this paradigm, a system is irradiated by a coherent laser pulse that dresses the electronic states, potentially changing their properties. The process allows modifying band dispersions, turning insulators into conductors and vice-versa, altering the crystal symmetry, and tuning the system's topology [2,3,9,15,20–32].

One of the most studied materials for light-induced band engineering is graphene, which in the absence of driving is a two-dimensional (2D) Dirac semimetal with band touching at the K and K' points in the Brillouin zone (BZ). The degeneracies can be lifted when driving the system with circularly-polarized light, generating diverse topological phases[15,18,22,23,26]. The effect is attributed to the breaking of time-reversal symmetry (TRS). The gap opening in light-driven graphene has not yet been observed in angle-resolved photoemission spectroscopy (ARPES) due to various possible experimental limitations[32,33,35], but hybridization gaps have been seen in other systems[2,36]. On the other hand, topologically-trivial gap opening in graphene also occurs without breaking TRS if inversion symmetry is lifted[37], or strain is introduced (moving the Dirac nodes until oppositely-charge nodes annihilate to open a gap)[38,39]. It was also shown in optical lattices that by shaking the lattice one can move the Dirac nodes along high-symmetry axes until they merge[40,41], with analogous phenomena occurring in the very intense high frequency driven regimes[42].

Here we show that in the strong-field and low frequency regime, a linearly-polarized monochromatic field can move the Dirac nodes in the BZ by a substantial amount, and the movement's direction is fully controlled by the laser polarization. Effectively, this opens a large pseudo-gap at the original position of the Dirac nodes. We analytically show that the physical mechanism for the effect relies on band nonlinearities, and therefore doesn't appear in the simplest linearized low-energy model of Dirac bands. We validate these results with time-dependent density functional theory (TDDFT) calculations of ARPES spectra. Lastly, we show that this physical mechanism is general and allows versatile control of band engineering in a wide range of materials. As examples, we demonstrate control over the position of the valley minima and valleytronics in hexagonal-Boron-Nitride (allowing valleytronics control in transition-metal-dichalcogenides)[43–46], splitting and moving charge-II Weyl cones[47], and merging Dirac nodes in three-dimensional (3D) Dirac semimetals[48,49].

We begin by analyzing a graphene system with a two-band tight-binding (TB) model with 5$^{th}$-order nearest-neighbor (NN) terms[50]. In the basis of creation/annihilation operators on the A/B sublattice sites of the honeycomb lattice, the field-free Hamiltonian is:

$$\widehat{H}_0 = \begin{pmatrix} t_2 f_2(\mathbf{k}) + t_5 f_5(\mathbf{k}) & t_1 f_1(\mathbf{k}) + t_3 f_3(\mathbf{k}) + t_4 f_4(\mathbf{k}) \\ t_1 f_1^*(\mathbf{k}) + t_3 f_3^*(\mathbf{k}) + t_4 f_4^*(\mathbf{k}) & t_2 f_2(\mathbf{k}) + t_5 f_5(\mathbf{k}) \end{pmatrix} + (3t_2 - 6t_5)\widehat{\sigma_0} \quad (1)$$

where $\widehat{\sigma_0}$ is the identity matrix, $t_i$ are hopping amplitudes to the $i$'th NN site, and $f_i(\mathbf{k})$ are structure factors (see supplementary information (SI) section I). The second term in Eq. (1) conveniently sets the top of the valence band to zero energy, while the first term represents the various hopping processes. Note that $\widehat{H}_0$ inherently does not include $\widehat{\sigma}_z$, setting the gap to zero and resulting in Dirac cones with local linear dispersion in the K/K' points. The eigenvalues of $\widehat{H}_0$, denoted as $\epsilon_\pm(\mathbf{k})$, are obtained analytically. The hopping amplitudes are fitted through least-squares such that $\epsilon_\pm(\mathbf{k})$ match bands obtained from density functional theory (DFT) calculations performed using octopus code[51–53] within the local density approximation (see SI sections II and III). Notably, the TB model provides very good bands around the K/K' valleys (the main region of interest), but fails around the Γ-point.

$\widehat{H}_0$ is coupled to an external laser by Peierls substitution, yielding $\widehat{H}(\mathbf{k}, t) = \widehat{H}_0\left(\mathbf{k} - \frac{1}{c}\mathbf{A}(t)\right)$, where $\mathbf{A}(t) = \frac{cE_0}{\omega}\sin(\omega t)\hat{\mathbf{e}}$ is light's vector potential within the dipole approximation. Here $E_0$ is the field amplitude, $\omega$ is the driving frequency, $c$ is the speed of light, and $\hat{\mathbf{e}}$ is a polarization vector. From this time-periodic Hamiltonian we obtain the Floquet Hamiltonian in the basis of harmonic functions of $\omega$ with the sub-blocks:

$$\widehat{H}_F^{n,m}(\mathbf{k}) = \delta_{n,m} n\omega \widehat{\sigma_0} + \frac{\omega}{2\pi}\int_0^{2\pi/\omega} dt\, \widehat{H}(\mathbf{k}, t)e^{i(n-m)\omega t} \quad (2)$$

where $|n - m|$ is the photon channel order, and the integrals in Eq. (2) are solved numerically. $\widehat{H}_F(\mathbf{k})$ is then exactly diagonalized, and the eigen-energies are corrected by their photon-channel index. The resulting Floquet quasi-energy valence and conduction bands, $\epsilon_\pm^F(\mathbf{k})$, are taken as the bands that converge to the field-free bands for $E_0 \to 0$.

Our main interest is the position of the Dirac nodes in the driven system. Since in graphene the Dirac nodes host a nonzero Berry phase[54–56], they cannot be removed by a linearly-polarized laser field (that does not break inversion or TRS[57]) unless oppositely-charged nodes merge[58]. However, we can still track the nodes' movements with respect to the laser driving. In order to simplify the analysis, we initially ask whether a Floquet quasi-energy gap can open in the original positions of the Dirac nodes at K/K', defined as $E_g^F = \epsilon_+^F(\mathbf{K}) - \epsilon_-^F(\mathbf{K})$. If a gap opens, the linearly-dispersing nodes have moved (note that we will later analyze directly the movement of the nodes). We analyze the Floquet propagator, $\widehat{U}_F(\mathbf{k}) = \exp\left(-i\int_0^{2\pi/\omega} \widehat{H}(\mathbf{k}, t)\right)$, and use atomic units unless stated otherwise. $\widehat{U}_F$ describes time propagation over one laser cycle, and taking the logarithm of its eigenvalues is formally equivalent to diagonalizing the Floquet Hamiltonian[27]. The propagator can be represented by a time-independent effective Hamiltonian[26,59], $\widehat{U}_F(\mathbf{k}) = \exp\left(-i\frac{2\pi}{\omega}\widehat{H}_{eff}(\mathbf{k})\right)$, where $\widehat{H}_{eff}$ comprises a Magnus series expansion:

$$\widehat{H}_{eff}(\mathbf{k}) = \widehat{H}_1(\mathbf{k}) + \widehat{H}_2(\mathbf{k}) + \widehat{H}_3(\mathbf{k}) + \cdots$$
$$\widehat{H}_1(\mathbf{k}) = \frac{\omega}{2\pi}\int_0^{\frac{2\pi}{\omega}} dt \widehat{H}(\mathbf{k}, t),\ \widehat{H}_2(\mathbf{k}) = \frac{-i\omega}{4\pi}\int_0^{\frac{2\pi}{\omega}} dt \int_0^t dt' [\widehat{H}(\mathbf{k}, t), \widehat{H}(\mathbf{k}, t')], \cdots \quad (3)$$

In this representation $\widehat{H}_1$ acts as a direct time-averaged Hamiltonian, and higher orders capture effects due to the Hamiltonian not commuting with itself at different times. This notation is especially appealing for analyzing gap openings in graphene, because $\widehat{H}(\mathbf{k}, t)$ does not include $\widehat{\sigma}_z$ terms; hence, $\widehat{H}_1$ purely comprises $\widehat{\sigma_0}, \widehat{\sigma_x}$, and $\widehat{\sigma_y}$ terms, and $\widehat{H}_2$ purely comprises gap-opening $\widehat{\sigma}_z$ terms. The next orders follow such that only even-order terms in the Magnus expansion allow potential gap openings at K/K'.

The main question of interest is under which conditions the even-order terms vanish. Let us first prove that for the perfectly-linear low-energy Dirac Hamiltonian, fields that do not break TRS cannot open a gap at K. For this, we take the first-order expansion of $\widehat{H}(\mathbf{k},t)$ around K, $\widehat{H}_D(\mathbf{k}) = v_f(\Delta k_x \widehat{\sigma_x} + \Delta k_y \widehat{\sigma_y})$, where $\Delta \mathbf{k}$ is the momenta away from K and $v_f$ is the Fermi velocity. Coupling $\widehat{H}_D(\mathbf{k})$ to an external laser field provides a time-periodic Hamiltonian that is inserted in the Magnus expansion. Due to the linearity of the Dirac Hamiltonian, we obtain $\widehat{H_1} = \widehat{H}_D$ (because $\int_0^{2\pi/\omega} \mathbf{A}(t)=0$). Thus, for the Dirac Hamiltonian only higher order terms can alter the band structure. For $\widehat{H_2}$ we find:

$$\widehat{H}_2(\Delta \mathbf{k}) = \frac{v_f^2}{c^2} \frac{\omega}{2\pi} \widehat{\sigma}_z \int_0^{\frac{2\pi}{\omega}} dt \int_0^t dt' \begin{bmatrix} \Delta k_x \left(A_y(t') - A_y(t)\right) + \Delta k_y \left(A_x(t) - A_x(t')\right) \\ + \left(A_x(t)A_y(t') - A_x(t')A_y(t)\right) \end{bmatrix} \quad (4)$$

There are three main terms inside the integral in Eq. (4): the first two in the top row vanish at K ($\Delta\mathbf{k}=0$). The third term is $k$-independent, and the only one that survives at K. This term clearly vanishes for linear driving, since then one of the laser polarization components is zero. We now show that it also vanishes for any TRS field. First, we separate the double integral in the third term in Eq. (4):

$$\int_0^{\frac{2\pi}{\omega}} dt' \int_0^t dt \left(A_x(t)A_y(t') - A_x(t')A_y(t)\right) = \int_0^{\frac{2\pi}{\omega}} dt A_x(t) \int_0^t dt' A_y(t') - \int_0^{\frac{2\pi}{\omega}} dt A_y(t) \int_0^t dt' A_x(t') \quad (5)$$

Next, without loss of generality we represent $\mathbf{A}(t)$ with a pure harmonic sine series, $\mathbf{A}(t) = \sum \mathbf{a}_m \sin(m\omega t)$, such that the electric field is given by a pure cosine series, $\mathbf{E}(t) = -\frac{1}{c}\partial_t \mathbf{A}(t) = -\omega \sum \mathbf{a}_m m\cos(m\omega t)$, inherently respecting TRS ($\mathbf{E}(t) = \mathbf{E}(-t)$). Plugging these into Eq. (5), we note that one polarization component of $\mathbf{A}(t)$ is always integrated over in the $t'$ integral, giving a time-even function, while the other component remains time-odd. The second temporal integral over $t$ then vanishes since it integrates a time-odd function. Thus, $\widehat{H_2} = 0$ in the Dirac Hamiltonian for any TRS drive. In the SI (section VI) we generalize this proof to all even orders of the Magnus expansion. Overall, a Floquet pseudo-gap cannot open in the low-energy Dirac Hamiltonian driven by a TRS field. This is a well-established result that has been shown with other methodologies. It is however potentially misleading, because it seemingly pinpoints the physical reason a gap does not open at K to the presence of TRS. Contrarily, we argue that the physical origin of the effect is the linearity of the Hamiltonian (and similarly, Weyl[13] or other linearly-dispersing systems[7]). Indeed, if one repeats the analysis for a field-free parabolic Hamiltonian of the form $H(\mathbf{k}) = v(\Delta k_x^2 \widehat{\sigma_x} + \Delta k_y^2 \widehat{\sigma_y})$, the proof no longer holds regardless of TRS. One can verify that in that case a Floquet gap does open, even though the Hamiltonian is spherically-symmetric and in a low-energy continuum form.

For completeness, we repeat the analysis for the TB Hamiltonian at K, keeping only up to 2$^{nd}$-order NN terms and employing a linearly-polarized drive along $k_y$ (respecting TRS). $\widehat{H_2}(\mathbf{K})$ takes the form:

$$\widehat{H}_2(\mathbf{K}) = 2t_1^2 \widehat{\sigma}_z \frac{\omega}{2\pi} \int_0^{\frac{2\pi}{\omega}} dt \int_0^t dt' g(t,t') \quad (6)$$

where the function under the double integral is comprised of nested trigonometric functions (see SI section VI) and cannot be analytically integrated. Still, Eq. (6) can be evaluated numerically, and we have found that generally $\widehat{H}_2(\mathbf{K}) \neq 0$. The SI (section VI) presents exemplary results for the size of $\widehat{H}_2(\mathbf{K})$ vs. the laser amplitude and wavelength, showing power-law like scalings. We also note some analytical intuition arises from this analysis, e.g. that the gap at K should scale parabolically with $t_1$, and be independent of $t_2$ (as well as $t_5$) because they only couple to $\widehat{\sigma_0}$ terms that commute. However, the size of the gap and its scaling with the laser parameters is not expected to correspond well with the size of $\widehat{H}_2(\mathbf{K})$, because in practical conditions higher order terms in the magnus expansion cannot be neglected[60,61]. Moreover, in the 5$^{th}$-NN TB Hamiltonian the $t_1$ hopping term interferes with higher-order terms, leading to more complex dynamics (see SI section VI). Nonetheless, even if not quantitative, this analysis establishes the gap opening at K and its physical origin – if $\widehat{H}_2(\mathbf{K}) \neq 0$, higher order terms will also be nonzero, and there is no general symmetry constraint that causes their summation to vanish. In the SI (section VI) we provide thorough exact numerical investigations of the size of the pseudo-gap – it indeed scales parabolically with $t_1$, and does not scale with $t_2$, corroborating the analytical analysis. We generally found that the gap at K can be very substantial (up to 0.5eV). Practically, we recall that this pseudo-gap means that the Dirac nodes moved elsewhere, where a larger gap suggests the positions have moved further away from K/K'.

Before moving further, it's worth highlighting some noteworthy points: (i) If $E_0/\omega$~1, the Magnus series can converge very slowly, or even diverge, but it's still valid for determining if a gap opens at K. (ii) The gap at K (and Dirac nodes movement) arises from band nonlinearity in the field-free Hamiltonian away from K, and it vanishes in the limit where the low-energy Dirac Hamiltonian becomes valid. However, simply evaluating the Hamiltonian in the vicinity of K does not guarantee that the low-energy expansion around it is valid; rather, $E_0/\omega$ must be sufficiently small. This condition breaks if $E_0$ is large, or $\omega$ is small, as obtained in the strong-field limit, and it's already broken in often employed conditions for observing Floquet sidebands (powers of ~$10^{11}$ W/cm$^2$ and wavelengths ~1600nm open a pseudo-gap of ~50meV and moves the Dirac node ~0.3% of the BZ along $k_x$). (iii) The Dirac node motion strongly depends on the laser orientation, since that greatly changes the Magnus expansion. For instance, for a laser polarized along $k_x$ we obtain $g(t,t')$=0, and only even terms beyond the 4$^{th}$-order in the Magnus expansion are nonzero.

Having established this result, we numerically investigate its dependence on the laser parameters. When graphene is driven along high symmetry axes (along Γ-*M* or Γ-*K*), we find that the Dirac nodes only move along the $k_x$ axis (similarly to shaken optical lattices[40]). Figures 1(a,b) present the distance of the out-of-equilibrium Dirac nodes from K (Δ*K*) *vs.* laser power and wavelength, which can be quite substantial, and up to ~10% of the BZ in reasonable experimental conditions. In more extreme cases, oppositely-charged Dirac nodes can even merge. We determined that this process requires laser powers of ~$10^{13}$ W/cm$^2$ (at 1600nm driving wavelength along the *x*-axis), although this value is qualitative because it depends on the details of the TB model around Γ, where it fails. This critical power is slightly higher than graphene's damage threshold, implying that linearly-polarized driving cannot open a proper gap.

Slightly different results are observed for *y*-polarized driving, where the Dirac nodes move in the opposite direction (see Figs. 1(a,b)). Interestingly, here the Dirac node at K(K') interact with the hybridized Floquet sidebands (replicas of K'(K)), until they gradually merge and open a gap for a certain critical pump wavelength and power. At that point, another gapless sideband enters the region. Such dynamics continue for longer wavelength (or higher intensities), where more sidebands enter the region around K/K'. The effect is similar to phenomena observed in other driving conditions with Dirac point spawning[26,37], but seems distinct to very long wavelength driving at which it also becomes difficult to distinguish between Floquet replicas and the original Dirac points (see SI section VI). Thus, measuring the position of the Dirac nodes with respect to the driving parameters could potentially probe additional information about the system such as band hybridization.

Figures 1(c,d) plots the position of the Dirac nodes in the driven system *vs.* the laser polarization axis (both angle, $\theta_K$ – the angle of the shifted Dirac point around its original position, and distance, $R_K$ – the distances of the shifted Dirac point from its original position, see illustrated trajectories in the inset of Fig. 1(d)). As the laser polarization rotates, the Dirac nodes smoothly rotate (with a trigonal pattern) around their equilibrium positions in correspondence. This verifies that a single monochromatic linearly-polarized laser can arbitrarily place the Dirac nodes in the BZ.

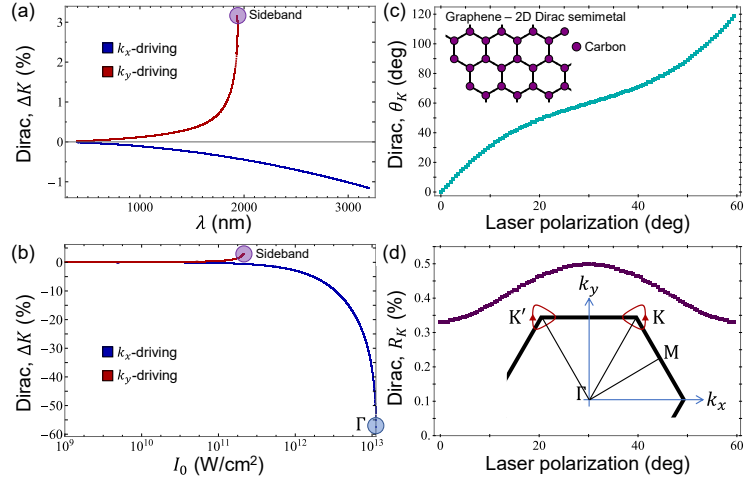

**FIG. 1.** Dirac node motion in graphene driven by linearly-polarized light. (a) Floquet-Dirac node distance from original K-point along the $k_x$-axis *vs.* driving wavelength for a power of $10^{11}$ W/cm$^2$, and driving along *x* and *y* axes. (b) Same as (a), but *vs.* driving power for a wavelength of 1600nm. The highlighted point signifies a Dirac nodes merger event (blue), and merger events with Floquet replicas (purple). (c) Angle of the Dirac node (with respect to the $k_x$-axis) *vs.* the driving polarization angle (with respect to the *x*-axis), for power of $10^{11}$ W/cm$^2$ and wavelength of 1600nm. (d) Same as (c), but presenting the radial distance of the Dirac node from its equilibrium position at K. Inset in (d) shows the trajectory (red) of the Dirac nodes around K/K' in the BZ as the laser polarization rotates (size enhanced for clarity). Inset in (c) shows the graphene lattice.

We next show that these results are not specific to graphene, or even to linearly-dispersing systems – band nonlinearity inherently exists in all periodic systems regardless of their low-energy local structure. First, we perform similar calculations in monolayer hexagonal-Boron-Nitride (hBN) (see SI section V and ref. 62 for details). Figures 2(a,b) shows that the position of the valley minima (defined as the minimal optical gap points in the BZ) move around with the laser drive and rotate around their equilibrium position by few percent. This provides a potential path to optically tune valley selectivity (also in transition-metal-dichalcogenides) without circular driving, because the local orbital character around the minima point differs from that at K/K', and the valley minima can be arbitrarily shifted away (whereas with circular driving it is fixed due to rotational symmetry). This is especially clear if one considers that valley optical selection rules can be explicitly derived only at K/K' points, while the Bloch states have mixed character in their vicinity[45,46]. Second, we perform similar calculations in the 3D Dirac semimetal, Na$_3$Bi[48,49] (see SI section V). Crucially, in Na$_3$Bi even the low-energy Hamiltonian contains large nonlinearities at the Dirac nodes, because they arise from a crossing of two parabolic bands. Figure 2(c) shows that linearly-polarized laser driving can move the Dirac nodes just as in graphene. The two nodes merge at laser powers of ~$10^{11}$ W/cm$^2$ (at 1600nm), which is within experimental feasibility. Physically, this merging is possible in Na$_3$Bi because the nodes are initially relatively close to each other. However, this predicted critical power might slightly differ in the realistic system due to the validity of the low energy Hamiltonian around Γ. Third, we calculate the Floquet quasi-energy bands for linearly-polarized driven SrSi$_2$, which is a Charge-II Weyl semimetal with parabolically-dispersing Weyl cones[47]. Due to the parabolic dispersion, the system is inherently nonlinear (see SI section V). We find that the laser splits the Charge-II Weyl cone into two Charge-I linearly-dispersing cones. Figure 2(d) shows that as the driving power increases, the new charge-I cones move further apart. We have found that their motion can be fully controlled within the *xy*-plane in which the field-free bands are parabolic (following the laser polarization). Along the $k_z$-axis on the other hand, the electronic bands are linearly-dispersing and no driving parameters can move the Weyl cones. This highlights that the physical mechanism relies on band nonlinearity.

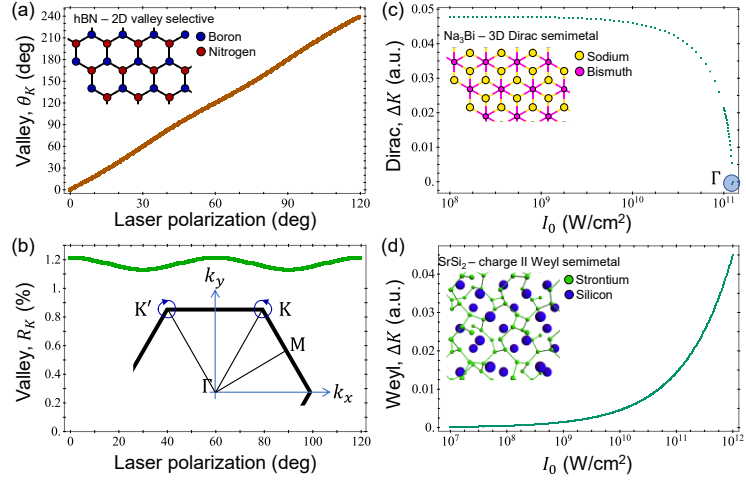

**FIG. 2.** Results in other material systems. (a) Angle of the valley minima point in hBN with the same notations and conditions as in Fig. 1. (b) Same as (a), but presenting the distance of the Dirac node from its equilibrium position at K. (c) Floquet-Dirac node motion in Na$_3$Bi – Dirac node distance from the original position *vs.* power, for a wavelength of 1600nm and *x*-axis polarization. The highlighted point signifies a Dirac node merger event. (d) Charge-II Weyl cone splitting, and Charge-I Weyl cone motion, in SrSi$_2$ – Floquet-Weyl cone distance from its original position *vs.* driving power for similar laser conditions as (c). Inset in (b) shows the trajectory (blue) of the valley minima around K/K' in the BZ as the laser polarization rotates (size enhanced for clarity). Insets in (a), (c), and (d), show the hBN, Na$_3$Bi, and SrSi$_2$, lattice structures, respectively.

Lastly, to further establish the model results, we perform *ab-initio* TDDFT calculations of ARPES in light-driven graphene. The methodology follows ref. 63, but with artificially doping the conduction band to make the ARPES signals from it more visible. All details of these calculations are delegated to the SI section IV (see also refs. 51–53,64–67). Figure 3 presents the resulting spectra along $k_x$ and $k_y$ axes overlayed with the quasi-energy bands obtained from the model, which agree remarkably well – a large gap of ~0.18eV opens at the original K point (seen when plotting along $k_y$), and the Dirac nodes shifts by ~1.15% of the BZ along $k_x$. Note that the use of the Dirac Hamiltonian in this case completely fails in describing the spectra because it fixes the Dirac nodes to K/K'. We further emphasize that even though intense pumping is required to observe these phenomena in ARPES, intensities of up to $4\times10^{10}$ W/cm$^2$ are already achievable[68], and work is underway to allow even more intense pumping[69]. Moreover, by utilizing longer wavelength pumps (e.g. in the THz regime[70]), weaker peak powers can be used to observe similar phenomena (see SI section VI). Regardless, even weaker signals of Dirac point motion could be extracted from experimental spectra by subtracting the field-free backgrounds or utilizing only their asymmetric part. Furthermore, as the motion is polarization dependent, spectra obtained at different polarizations will help distil the signal. Therefore, these predictions should be experimentally accessible with current technology.

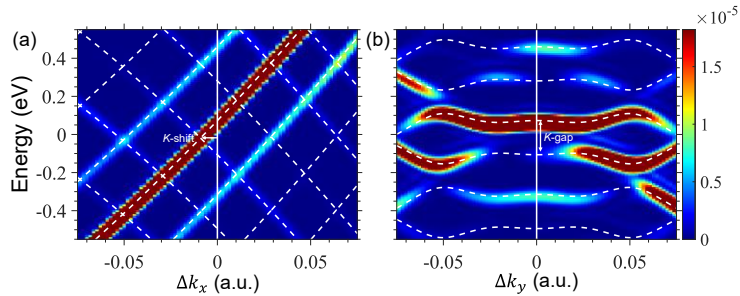

**FIG. 3.** *Ab-initio* TDDFT calculations of ARPES from light-driven graphene for linearly-polarized driving along $k_x$ at 3200nm and $10^{11}$ W/cm$^2$. The spectrum is plotted along $k_y$ (a), and $k_x$ (b), in the region of K, and is saturated for clarity. The overlayed dashed lines denote the Floquet quasi-energy bands obtained from the model in the same driving conditions. Arrows indicate shifting of the Dirac node and opening a gap at K.

To conclude, we investigated several material systems irradiated by intense low-frequency linearly-polarized lasers. For Dirac linearly-dispersing systems, we showed that the laser moves the Dirac nodes away from their initial position. This motion is substantial and can be fully controlled by changing the laser parameters (intensity, wavelength, polarization). The effect was analytically shown to originate from band nonlinearities, highlighting the importance of the employed model. Consequently, our results emphasize the obvious, yet sometimes overlooked feature, that low-energy Hamiltonians fail when driven by sufficiently intense or long-wavelength lasers. We further validated the generality of the physical mechanism with extensive additional calculations, showing that linearly-polarized driving can: (i) control the positions of valley minima in valley-selective materials (tuning valleytronics), (ii) merge Dirac nodes in 3D Dirac semimetals, and (iii) split high-order Weyl cones and control the positions of the resulting linearly-dispersing cones. We confirmed the model results with *ab-initio* TDDFT calculations and outlined an ARPES set-up able to test our predictions.

The present findings should help guide future experiments and theory of Floquet band engineering; and in particular, to benefit from electronic band structure nonlinearities to tailor material properties. Our results also emphasize the importance of the full BZ and band structure away from the minimal gap points in strong-field physics processes in solids, such as high harmonic generation[71–73], photogalvanic effects[74–76], magneto-optical effects[77,78], and more. This is especially relevant in quantum materials and systems with topological or linearly-dispersing bands[70,79–83], motivating development of *ab-initio* methodologies. We expect that the movement of the high-symmetry points in the BZ will imprint additional characteristics not only directly in ARPES, but also for linear and nonlinear optical responses such as transient absorption spectra and high harmonic generation, which should motivate future work.

**SUPPORTING INFORMATION.** Technical details of the tight-binding model. Technical details about the DFT calculations and fitting procedures of the tight binding hooping amplitudes. Technical details of the TDDFT calculations and ARPES calculations. Technical details of the Floquet calculations in material systems other than graphene. Extended proof that all higher-order even Magnus expansion terms vanish in the Dirac Hamiltonian driven by time-reversal symmetric light. Extended numerical investigation of the pseudo-gap opening in graphene and its scaling with laser parameters and tight-binding parameters. Additional results of tr-ARPES in graphene for other laser parameters.

**ACKNOWLEDGEMENTS.** This work was supported by the Cluster of Excellence Advanced Imaging of Matter (AIM) – EXC 2056 - project ID 390715994, SFB-925 "Light induced dynamics and control of correlated quantum systems", project 170620586 of the Deutsche Forschungsgemeinschaft (DFG), Grupos Consolidados (IT1453-22), and the Max Planck-New York City Center for Non-Equilibrium Quantum Phenomena. The Flatiron Institute is a division of the Simons Foundation. O.N. gratefully acknowledges the generous support of a Schmidt Science Fellowship.

# Supplementary information:
# Band nonlinearity-enabled manipulation of Dirac nodes, Weyl cones, and valleytronics with intense linearly polarized light

**Ofer Neufeld[1,*], Hannes Hübener[1], Gregor Jotzu[1], Umberto De Giovannini[1,2], Angel Rubio[1,3,*]**

[1]Max Planck Institute for the Structure and Dynamics of Matter and Center for Free-electron Laser Science, Hamburg 22761, Germany.
[2]Università degli Studi di Palermo, Dipartimento di Fisica e Chimica—Emilio Segrè, Palermo I-90123, Italy.
[3]Center for Computational Quantum Physics (CCQ), The Flatiron Institute, New York 10010, USA.
*Corresponding author E-mails: oneufeld@schmidtsciencefellows.org, angel.rubio@mpsd.mpg.de.

## I. GRAPHENE TIGHT-BINDING HAMILTONIAN DETAILS

We provide here additional technical details on the TB Hamiltonian employed throughout the simulations presented in the main text. The lattice primitive vectors are given as $\mathbf{a}_1 = a_0\hat{\mathbf{x}}$, $\mathbf{a}_2 = a_0(-\hat{\mathbf{x}}/2 + \sqrt{3}\hat{\mathbf{y}}/2)$, with the graphene experimental lattice parameter $a_0$=2.456Å. The NN hopping vectors on this lattice are provided in Table S1 below with the notation $\mathbf{v}_{i,j}$, where $i$ is the order of the hopping process (e.g. $i$=3 is 3$^{rd}$ NN hopping), and $j$ is the index of the vector (there are either three or six vectors for a given hopping process, see Fig. S1 for illustration). The resulting structure factors $f_i(\mathbf{k})$ are given as:

$$f_m(\mathbf{k}) = \sum_n \exp\{i\mathbf{k}\cdot\mathbf{v}_{m,n}\} \qquad (S1)$$

where the sum runs over all existing $n$'s for that particular order of hopping.

**Table S1** – NN hopping vectors, $\mathbf{v}_{i,j}$, given in basis of real space vectors in 2D.

| NN order | Hopping vectors | | | | | |
|---|---|---|---|---|---|---|
| 1 | $a_0\left\{0,\frac{1}{\sqrt{3}}\right\}$ | $a_0\left\{-\frac{1}{2},-\frac{1}{2\sqrt{3}}\right\}$ | $a_0\left\{\frac{1}{2},-\frac{1}{2\sqrt{3}}\right\}$ | | | |
| 2 | $a_0\left\{-\frac{1}{2},-\frac{\sqrt{3}}{2}\right\}$ | $a_0\left\{\frac{1}{2},\frac{\sqrt{3}}{2}\right\}$ | $a_0\{-1,0\}$ | $a_0\{1,0\}$ | $a_0\left\{-\frac{1}{2},\frac{\sqrt{3}}{2}\right\}$ | $a_0\left\{\frac{1}{2},-\frac{\sqrt{3}}{2}\right\}$ |
| 3 | $a_0\left\{1,\frac{1}{\sqrt{3}}\right\}$ | $a_0\left\{-1,\frac{1}{\sqrt{3}}\right\}$ | $a_0\left\{0,-\frac{2}{\sqrt{3}}\right\}$ | | | |
| 4 | $a_0\left\{-1,-\frac{2}{\sqrt{3}}\right\}$ | $a_0\left\{\frac{1}{2},\frac{5}{2\sqrt{3}}\right\}$ | $a_0\left\{-\frac{3}{2},-\frac{1}{2\sqrt{3}}\right\}$ | $a_0\left\{\frac{3}{2},-\frac{1}{2\sqrt{3}}\right\}$ | $a_0\left\{-\frac{1}{2},\frac{5}{2\sqrt{3}}\right\}$ | $a_0\left\{1,-\frac{2}{\sqrt{3}}\right\}$ |
| 5 | $a_0\left\{-\frac{3}{2},-\frac{\sqrt{3}}{2}\right\}$ | $a_0\left\{\frac{3}{2},\frac{\sqrt{3}}{2}\right\}$ | $a_0\left\{\frac{3}{2},-\frac{\sqrt{3}}{2}\right\}$ | $a_0\left\{-\frac{3}{2},\frac{\sqrt{3}}{2}\right\}$ | $a_0\{0,\sqrt{3}\}$ | $a_0\{0,-\sqrt{3}\}$ |

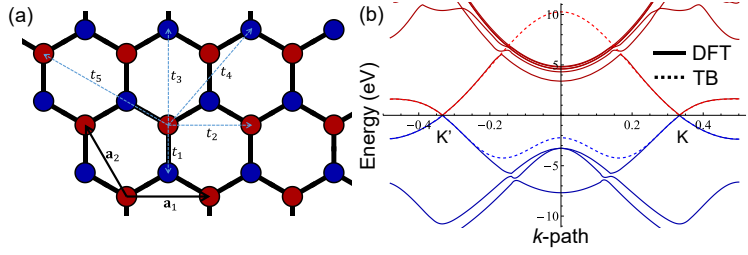

**FIG. S1.** System illustration. (a) Schematic graphene lattice model with NN hopping terms. Red (blue) denote A (B) sublattice sites, and arrows indicate the different hopping processes and lattice vectors. (b) DFT obtained bands for graphene along a path traversing the BZ from Γ through K in fractional coordinates (solid), compared to the TB model (dashed). Blue and red denote occupied and unoccupied bands, respectively.

## II. GROUND-STATE DFT CALCULATIONS

We provide here technical details for performed DFT calculations, which were also employed for obtaining initial states for the TDDFT simulations outlined below (for calculating the ARPES spectra presented in the main text). All DFT calculations were performed with Octopus code[1–3] in a real-space grid representation. The grid was represented on the non-orthogonal primitive unit cell of graphene with equidistant spacings of 0.38 Bohr along the lattice vectors, periodic boundary conditions in the monolayer plane (*xy* plane), and finite boundary conditions along the *z*-axis (where the total length of the *z*-axis was converged at 110 Bohr). We used the experimental lattice parameter of graphene, $a_0$=2.456Å. A discrete *k*-grid was converged at a Γ-centered 12x12x1 grid for representing the electron density, but a much finer mesh of 36x36x1 *k*-points was employed for outputting the band structure for the fitting procedures described below (presented in Fig. S1). Calculations were performed within the local density approximation (LDA) for the exchange-correlation (XC) functional, and while neglecting spin degrees of freedom and spin-orbit coupling. We employed norm-conserving pseudopotentials for describing core states of Carbon[4]. The Kohn-Sham (KS) equations were solved to a strict self-consistency tolerance of $10^{-9}$ Hartree per unit cell.

## III. HOPPING AMPLITUDES FITTING PROCEDURE

From the ground-state DFT calculations we obtained the KS eigenvalues on a finite *k*-grid, $\epsilon_{KS,n}(\mathbf{k_i})$, where $n$ is the band index and $k_i$ is a grid point in the BZ. The KS valence ($n=4$) and conduction ($n=5$) band eigenvalues were then fitted to the TB model bands, $\epsilon_\pm(\mathbf{k})$, by employing a least-squares fitting procedure. We optimized the following target function:

$$M(t_1,t_2,t_3,t_4,t_5) = \sum_i \left|\epsilon_{KS,5}(\mathbf{k_i}) - \epsilon_+(\mathbf{k_i})\right| + \sum_i \left|\epsilon_{KS,4}(\mathbf{k_i}) - \epsilon_-(\mathbf{k_i})\right| \tag{S2}$$

where the sum included all discrete *k*-points in the BZ that upheld the condition $\frac{2\pi}{a_0}(k_{i,x} + \sqrt{3}k_{i,y}) \geq 0.5$. This condition essentially selects points within the K and K' valleys for the fitting procedure (removing points near Γ where the band dispersions invert), and further utilized TRS. The resolution of the *k*-grid used for fitting was 360x360x1, where the KS eigenvalues for points in-between the original *k*-grid were linearly-interpolated (with the original grid being 10-fold less dense, 36x36x1). This further guaranteed proper weights were given to the linear region around K and K'. The resulting fitted hopping amplitudes are: -2.0470, 0.4462, -0.0225, 0.1808, and 0.1021 eV, respectively, for $t_1$-$t_5$. Comparison between the band structures is presented in Fig. S1.

## IV. TDDFT-ARPES CALCULATIONS

We provide here the full details for the *ab-initio* TDDFT-ARPES calculations presented in the main text. We described the laser-induced electron dynamics within the KS-TDDFT framework, where the following KS equations of motion (in atomic units) were solved within the primitive unit-cell of the graphene lattice (with the additional vacuum spacing above and below the monolayer as discussed above):

$$i\partial_t |\varphi_{n,k}^{KS}(t)\rangle = \left(\frac{1}{2}\left(-i\mathbf{\nabla} + \frac{\mathbf{A}(t)}{c}\right)^2 + v_{KS}(\mathbf{r},t)\right)|\varphi_{n,k}^{KS}(t)\rangle \tag{S3}$$

where $|\varphi_{n,k}^{KS}(t)\rangle$ is the KS-Bloch state at $k$-point $k$ and band index $n$, $\mathbf{A}(t)$ is the total vector potential of all laser pulses interacting with matter within the dipole approximation, such that $-\partial_t \mathbf{A}(t) = c\mathbf{E}(t)$, $c$ is the speed of light in atomic units ($c\approx 137.036$). $v_{KS}(\mathbf{r},t)$ in Eq. (S3) is the time-dependent KS potential given by:

$$v_{KS}(\mathbf{r},t) = -\sum_I \frac{Z_I}{|\mathbf{R}_I - \mathbf{r}|} + \int d^3r' \frac{n(\mathbf{r}',t)}{|\mathbf{r}-\mathbf{r}'|} + v_{XC}[n(\mathbf{r},t)] \tag{S4}$$

where $Z_I$ is the charge of the $I$'th nuclei and $\mathbf{R}_I$ is its coordinate (describing the two carbon atoms in the graphene primitive unit cell), $v_{XC}$ is the XC potential that is a functional of $n(\mathbf{r},t) = \sum_{n,k}|\langle\mathbf{r}|\varphi_{n,k}^{KS}(t)\rangle|^2$, the time-dependent electron density (where we employed the adiabatic LDA approximation). Note that practically, the bare Coulomb interaction of electrons with the nuclei was replaced with non-local pseudopotentials (described above, assuming the frozen core approximation for core states of Carbon). We neglected coupling to phonons and assumed frozen ions.

These equations were propagated in time from the initial states obtained in ground state DFT calculations (with the 12x12x1 Γ-centered $k$-grid for describing $n(\mathbf{r},t)$), with a time step of 2.9 attoseconds. The main difference here compared to the calculations described above is that the ground state used for the ARPES calculations also involved an artificial doping of the graphene system to populate small electronic charges in the conduction band (making its contribution more visible in ARPES spectra). To this end, we added an additional $0.35\bar{e}$ charge to each unit cell, which was compensated for by an attractive potential arising from the following constructed classical charge density:

$$\rho_{dope}(\mathbf{r}) = N\exp\{-5z^2\} \tag{S5}$$

where $N = 0.02359$ is a normalization constant set such that the total charge from $\rho_{dope}$ integrates to $-0.35\bar{e}$ per unit cell. This guarantees that the system is neutral to avoid issues of charging with periodic boundary conditions. The classical electrostatic potential essentially binds the additional charge on the monolayer, making sure that it occupies graphene bands rather than continuum states. The entire procedure is roughly analogous to the experimental technique of adding a gate potential to dope the conduction band, and only slightly perturbs the overall electronic structure (the band structure is roughly unchanged with and without the doping procedure).

The time-dependent equations of motion were also solved on auxiliary $k$-grids along which the ARPES spectra was calculated. These grids traversed through the K point in the BZ and stretched along $k_y$, or $k_x$, with a total of 144 points, starting from $\mathbf{k} = \frac{2\pi}{3a_0}\left(1, \frac{2}{\sqrt{3}}\right)$ and ending at $\mathbf{k} = \frac{2\pi}{3a_0}\left(1, \frac{4}{\sqrt{3}}\right)$ for the grid that is parallel to $k_y$, and starting at $\mathbf{k} = \frac{2\pi}{a_0}\left(\frac{1}{3} - \frac{\sqrt{3}}{9}, \frac{1}{\sqrt{3}}\right)$ and ending at $\mathbf{k} = \frac{2\pi}{a_0}\left(\frac{1}{3} + \frac{\sqrt{3}}{9}, \frac{1}{\sqrt{3}}\right)$ for the grid that is parallel to $k_x$. However, the electron density was not contributed from this grid, but only from the 12x12x1 grid on which the ground state was calculated. During propagation we added a smooth complex absorbing potential (CAP) to avoid spurious reflection of electrons from the boundary. The CAP had a $\sin^2(z)$ shape that saturates to a height of -1 along the $z$-axis grid edges and a total width of 30 Bohr from both sides.

The employed vector potential $\mathbf{A}(t)$, comprised of two pieces – the pump pulse that induces a light-driven Floquet state (discussed in the main text), and an additional probe pulse that photoionizes electrons from the monolayer which can be detected in ARPES experiments. The resulting form is:

$$\mathbf{A}(t) = \mathbf{A}_{\text{pump}}(t) + \mathbf{A}_{\text{probe}}(t) \tag{S6}$$

, with

$$\begin{aligned}\mathbf{A}_{\text{pump}}(t) &= f(t)\frac{cE_0}{\omega}\sin(\omega t)\hat{\mathbf{x}} \\ \mathbf{A}_{\text{probe}}(t) &= f_{\text{xuv}}(t-t_0)\frac{cE_{\text{xuv}}}{\omega_{\text{xuv}}}\sin(\omega_{\text{xuv}}t)\hat{\mathbf{z}}\end{aligned} \tag{S7}$$

where $f(t)$ is a temporal envelope function taken to have the following 'super-sine' form[5]:

$$f(t) = \left(sin\left(\pi \frac{t}{T_p}\right)\right)^{\left(\frac{\left|\pi\left(\frac{t}{T_p}-\frac{1}{2}\right)\right|}{\sigma}\right)} \tag{S8}$$

where $\sigma$=0.75, $T_p$ is the duration of the laser pulse which was taken to be $T_p$=$16(2\pi/\omega)$. This form is roughly analogous to a super-gaussian, but where $f(t)$ starts and ends exactly at zero which is numerically convenient. The corresponding full-width-half-max (FWHM) of the pulse is ~85.5 femtoseconds for the chosen $\omega$, which corresponded to 3200nm wavelength driving, assuring the system enters a Floquet steady-state. The envelope function for the probe laser pulse was taken to have a similar form:

$$f_{xuv}(t) = \left(sin\left(\pi \frac{t}{T_{xuv}}\right)\right)^{\left(\frac{\left|\pi\left(\frac{t}{T_{xuv}}-\frac{1}{2}\right)\right|}{\sigma}\right)} \tag{S9}$$

where $T_{xuv}$ is the total duration of the probe pulse taken here as $T_{xuv} = 2000(2\pi/\omega_{xuv})$, which had a FWHM of 51.7 femtoseconds, such that the probe samples multiple cycles of the pump pulse and corresponds to probing Floquet-related physics. The photon energy of the probe was chosen as $\omega_{xuv}$=80 eV, and its intensity was taken as $2\times10^8$ W/cm$^2$ to only stimulate single-photon ionization. Both pulses were synchronized to overlap in time such that their peak powers coincided.

The ARPES spectra were calculated directly from the propagated KS states, and without additional fundamental assumptions, using the highly accurate and efficient surface-flux method T-SURFF[6,7]. The momentum-resolved flux of photoelectrons was recorded across a surface normal to the monolayer located at the onset of the CAP. The photoemission from all KS-Bloch states was coherently summed, producing $k$-resolved spectra along the auxiliary $k$-grid. The resulting spectra were smoothed with a moving mean filter and saturated to enhance the visibility of the emission lines.

## V. NUMERICAL DETAILS IN OTHER MATERIAL SYSTEMS
### 1. hBN CALCULATIONS

The TB model Hamiltonian employed for the hBN calculations was equivalent to that in eq. (1) in the main text, but with an added mass term of size $\Delta$ (opening a gap of size $\Delta$ at K and K')[8], which has the form:

$$\hat{H}_{0,hBN} = \hat{H}_{0,graphene} + \frac{\Delta}{2}(\hat{\sigma_0} + \hat{\sigma_z}) \tag{S10}$$

but where the hopping amplitudes $t_i$ all differ from those chosen from the graphene model. The hopping amplitudes chosen for hBN were fitted to ground-state DFT calculations for its band structure performed with a similar methodology to that described above for graphene (we used the experimental lattice constant of 2.52Å). The resulting employed hopping parameters were -2.1430, 0.3376, -0.0630, 0.1825, and 0.0928 eV, for $t_1$, $t_2$, $t_3$, $t_4$, and $t_5$, respectively, and $\Delta$ was set at 4.4269 eV (the DFT gap within LDA). The Floquet Hamiltonian and diagonalization procedures were the same as employed in graphene.

### 2. Na$_3$Bi CALCULATIONS

The model Hamiltonian employed for the three-dimensional Dirac semimetal Na$_3$Bi was adapted from the low energy continuum form of the TB expansion in ref.[9]. The resulting $4 \times 4$ field-free Hamiltonian has the from:

$$\hat{H}_{0,Na_3Bi} = \epsilon_0(\mathbf{k}) + \begin{pmatrix} M(\mathbf{k}) & A(k_x + ik_y) & & \\ A(k_x - ik_y) & -M(\mathbf{k}) & & \\ & & M(\mathbf{k}) & -A(k_x - ik_y) \\ & & -A(k_x + ik_y) & -M(\mathbf{k}) \end{pmatrix} \tag{S11}$$

where $\epsilon_0(\mathbf{k}) = C_0 + C_1 k_z^2 + C_2(k_x^2 + k_y^2)$, $M(\mathbf{k}) = M_0 - M_1 k_z^2 - M_2(k_x^2 + k_y^2)$, and all parameter values were taken as in ref.[9]. The Floquet Hamiltonian for the driven system was computed with the same approach used for the $2 \times 2$ system calculations, but where the additional $z$-axis and bands were also considered.

### 3. SrSi$_2$ CALCULATIONS

The model Hamiltonian employed for the three-dimensional charge-II Weyl semimetal SrSi$_2$ was taken from the low energy expansion in ref.[10] around the Weyl cones, including the spin-orbit interaction terms. The resulting $4 \times 4$ field-free Hamiltonian has the from:

$$\hat{H}_{0,SrSi_2} = \begin{pmatrix} v_t k_z & v(k_x - ik_y) & & \\ v(k_x + ik_y) & -v_t k_z & 2\Delta & \\ & 2\Delta & v_t k_z & v(k_x - ik_y) \\ & & v(k_x - ik_y) & -v_t k_z \end{pmatrix} \tag{S12}$$

where for the numerical calculations we employed arbitrary model parameters of $v=0.5$, $v_t=0.7$, and $\Delta=0.125$ a.u. The Floquet Hamiltonian for the driven system was computed in the same manner as the other systems.

## VI. ADDITIONAL RESULTS IN GRAPHENE
### 1. DIRAC HAMILTONIAN MAGNUS EXPANSION

In the main text we showed that the 2$^{nd}$-order term in the Magnus expansion of the Floquet Hamiltonian vanishes for the driven low-energy expanded Dirac Hamiltonian with any field that respects TRS. We present here the generalization of this proof to all higher order even terms. The $2n$'th-order term in the Magnus expansion generally comprises of temporal integrals of commutators of the form:

$$\hat{H}_{2n}(\mathbf{K}) \propto \int dt \int dt' \int dt'' \cdots \left[\hat{H}(\mathbf{k},t), \left[\hat{H}(\mathbf{k},t'), \left[\hat{H}(\mathbf{k},t''), \cdots \right]\right]\right] \tag{S13}$$

where there are $2n$ Hamiltonians appearing taking different time arguments, and $2n$ temporal integrals. All other permutations of the commutators also appear in summation, but it is enough to prove that one of them vanishes because the others are connected by permutations of the temporal arguments. Due to the linearity of the Dirac Hamiltonian, $\hat{H}_D(\mathbf{k})$, and the linearity of the Peierls substitution, the driven Hamiltonians that enter eq. (S10) can be separated into: $\hat{H}(\mathbf{k},t) = \hat{H}_D(\mathbf{k}) - \frac{1}{c}\hat{H}_D(\mathbf{A}(t))$. This results in different terms in the commutators that mix different orders of the vector potential. The terms can be sorted according to their power-law proportionality to the vector potential, ranging from $\propto (E_0)^0$, up to $\propto (E_0)^{2n}$. All of the terms that depend on $\mathbf{k}$ vanish when evaluated at K, and only one term in Eq. (S13) is $k$-independent, which is the term proportional to $(E_0)^{2n}$. That term arises from substituting in eq. (S13) $\hat{H}(\mathbf{k},t) \to -\frac{1}{c}\hat{H}_D(\mathbf{A}(t))$ for every temporal argument, such that it is necessarily proportional to:

$$\hat{H}_{2n}(\mathbf{K}) \propto \int dt \int dt' \int dt'' \cdots \left[\hat{\boldsymbol{\sigma}} \cdot \mathbf{A}(t), \left[\hat{\boldsymbol{\sigma}} \cdot \mathbf{A}(t'), \left[\hat{\boldsymbol{\sigma}} \cdot \mathbf{A}(t''), \cdots \right]\right]\right] \tag{S14}$$

where here $\hat{\boldsymbol{\sigma}}$ is the vector of Pauli matrices. Since only $\hat{\sigma}_x$ and $\hat{\sigma}_y$ terms exist in each $\hat{H}_D$, after $2n$ commutations we are left with one $\hat{\sigma}_z$ term, which is proportional to:

$$\hat{H}_{2n}(\mathbf{K}) \propto \hat{\sigma}_z \int dt \int dt' \int dt'' \cdots \sum_{i,j,p} A_i(t) A_j(t') A_p(t'') \cdots \tag{S15}$$

where the indices $i, j, p \ldots$ refer to the cartesian components of the vector potential, which alternate with different permutations for different temporal arguments. Importantly, each of the sums in Eq. (S12) contains a product of $2n$ vector potential terms taking different temporal arguments. Consequently, the first integral can be separated out just like performed in the main text for the 2$^{nd}$-order term. Another important point is that regardless of the permutations of the cartesian components in the sum in eq. (S12) (which are not described here), every cartesian component of $\mathbf{A}(t)$ upholds TRS on its own. Thus, if the field upholds TRS, then $A_i(t)$ is time-odd for any $i$, and the first integral generates a pure time-even function. Each subsequent integral flips the parity of the resulting function, because each vector potential in the product takes a unique time argument. After $2n - 1$ integrals the resulting function (under the last integral) is necessarily time-odd, such that the final integral vanishes. This mathematical result occurs in any permutation of the cartesian components and time arguments in the sum in Eq. (S15), and in every permutation of time arguments in eq. (S13). Therefore, we have proven that if the drive respects TRS, all even-order terms in the Magnus expansion of the Floquet Hamiltonian at K vanish for the driven Dirac Hamiltonian, such that there are no gap-opening terms at K. This result relies on the linearity of the Hamiltonian, as discussed in the main text.

## 2. TB FLOQUET HAMILTONIAN GAP-OPENING TERMS

We explore here the scaling of the first gap-opening term in the Magnus expansion of the Floquet Hamiltonian for the driven TB model of graphene. Specifically, we numerically compute the integrals of the $g(t,t')$ function discussed in the main text for varying laser power and wavelength. The function takes the form:

$$g(t,t') = \sin\left(\frac{S(t,t')}{3}\right)\begin{bmatrix} \cos\left(\frac{S(t,t')}{3}\right) + \cos\left(\frac{4\pi}{3} + \frac{S(t,t')}{3}\right) + \cos\left(\frac{4\pi}{3} - \frac{S(t,t')}{3}\right) \\ -\cos(S(t,t')) - \cos\left(\frac{2\pi}{3} + S(t,-t')\right) - \cos\left(\frac{2\pi}{3} - S(t,-t')\right) \end{bmatrix} \tag{S16}$$

with $S(t,t') = \frac{\sqrt{3}a_0 E_0}{4\omega}(\sin(\omega t) - \sin(\omega t'))$. Fig. S2 presents the numerical results showing a cubic power-law scaling with the field amplitude, and 5$^{th}$ power-law scaling with wavelength. Notably, these do not agree with the full numerical results obtained by directly diagonalizing the Floquet Hamiltonian (Fig. S3 in the next SI section). The main reason is that the Magnus expansion converges extremely slowly, and even diverges for high power and long wavelength drives. Nonetheless, the gap term clearly converges to zero in the limit of weak driving, as expected.

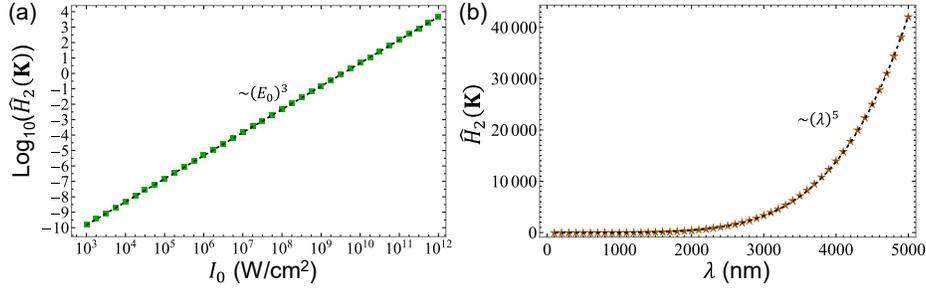

FIG. S2. (a) Scaling of the first gap-opening term in the Magnus expansion of the driven TB model (up to 2$^{nd}$ order TB terms) with wavelength for a driving power of $10^{11}$ W/cm$^2$. (b) Same as (a) but with field power and for a wavelength of 1600nm. Dashed black lines present fitted scaling laws.

We also present here the analytic expression for the 2$^{nd}$-order term in the Magnus expansion of the driven 5$^{th}$-order TB Hamiltonian (where the main text and Fig. S2 only discuss the driven 2$^{nd}$-order NN TB model while setting $t_3 = t_4 = t_5 = 0$). The resulting $\widehat{H}_2(\mathbf{k})$ for the same condition explored in the main text ($k_y$ driving by a linearly-polarized field) is:

$$\widehat{H}_2(\mathbf{K}) = i\hat{\sigma}_z \frac{\omega}{\pi} \int_0^{\frac{2\pi}{\omega}} dt \int_0^t dt' h(t,t') \tag{S17}$$

where the function under the integral is:

$$h(t,t') = \sin\left(\frac{\kappa \sin \omega t}{2}\right) \sin\left(\frac{\kappa \sin \omega t'}{2}\right) \begin{bmatrix} e^{i\kappa \frac{\sin \omega t - \sin \omega t'}{6}} p(t) p^*(t') \\ -e^{i\kappa \frac{\sin \omega t' - \sin \omega t}{6}} p^*(t) p(t') \end{bmatrix} \tag{S18}$$

where

$$p(t) = t_4 - t_1 + t_3 - i(t_3 - 2t_4)\sin(\kappa \sin \omega t) + t_3 \cos(\kappa \sin \omega t) \tag{S19}$$

, and with $\kappa = \frac{\sqrt{3}a_0 E_0}{2\omega}$. Numerical integration of eq. (S17) leads to very similar results to those presented in Fig. S2 for the 2$^{nd}$-order TB model (not presented), but the main difference here is that the interference between the different hopping amplitudes becomes apparent. In particular, different hopping terms contribute to separate parts of the integrand in eq. (S17). This further highlights the role of the full band structure in determining the gap opening at K.

## 3. K-GAP-OPENING

We numerically investigate here the size of the gap opening in graphene at K *vs.* the laser driving parameters, which complements the analysis presented in the main text for the position of the Dirac nodes in the BZ. Figure S3 shows the resulting exemplary Floquet gaps at K for various conditions, and their scaling with: (a) the NN hopping amplitude $t_1$, (b) the next-NN hopping amplitude $t_2$, (c) driving wavelength, and (d) driving power, for the case of a laser polarized along $k_y$. The gap indeed scales parabolically with $t_1$, and is independent of $t_2$, as expected from the analytical analysis. It exhibits an initial parabolic power-law scaling with the driving field amplitude (up until deviations appear from ~$5\times10^{10}$ W/cm$^2$ onwards), and an initial parabolic scaling with wavelength (until deviations arise from ~1000nm onwards). We also note that due to higher order NN hopping terms (that can have opposite signs), there can be non-trivial interference terms that cause gap closing and re-opening (e.g. *vs.* wavelength in Fig. S3(c) at ~2500nm, or *vs.* power in Fig. S3(d) at ~$6\times10^{11}$ W/cm$^2$). This arises when the vector potential term increases in magnitude and probes different regions in the bands (e.g. where their dispersion flips), and connects with the oscillations of the Dirac node around K discussed in the main text for *y*-polarized driving.

Figures S3(e) and (f) present the gap scaling with laser wavelength and power when the drive is polarized along $k_x$, which leads to slightly different behavior, but a similar initial scaling with field power and wavelength. The differences arise due to the distinct shape of the field-free bands along those lines. In particular, $k_x$ driving tends to open a larger gap that can be as high as ~0.5 eV in strong-field driving conditions. Note that in principle, such gap opening and re-closing dynamics could also be approximately described by effective time-independent Hamiltonians with the ground-state tunneling amplitudes modified by the laser (e.g. as in ref.[11]).

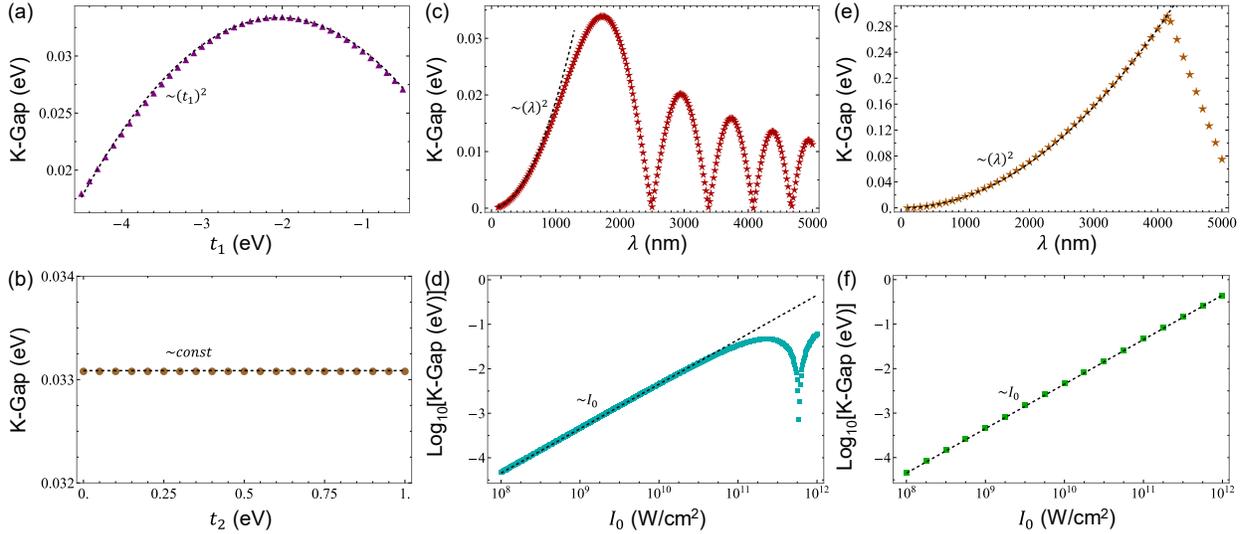

**FIG. S3.** Analysis of pseudo-gap opening in graphene driven by linearly-polarized light. (a) Floquet gap at K calculated numerically with the TB model for a laser power of $10^{11}$ W/cm$^2$ and 1600nm wavelength polarized along $k_y$ *vs.* the NN hopping amplitude, $t_1$. (b) Same as (a), but *vs.* 2$^{nd}$ NN hopping amplitude, $t_2$. (c) Same as (a) but for changing wavelength. (d) Same as (a) but for changing power, where the gap is shown in log scale. (e,f) Same as (c,d), but for driving along the $k_x$ axis. Dashed black lines in all cases present fitted scaling laws as discussed in the main text.

## 4. Interactions with Floquet replicas

We numerically explore the Floquet band structure *vs.* the pumping wavelength in the strong-field regime, and specifically, the Dirac point merger event with nearby sidebands indicated in Fig. 1(a) in the main text. Similar events occur when increasing the driving power as well. Fig. S4 shows the Floquet band structure while tuning the driving wavelength. The original Dirac point (which starts at K in the field-free case, and high frequency regime) slightly shifts right along the $k_x$-axis as the wavelength increases, until another sideband replica of K' approaches from the right. For a critical wavelength of ~1940 nm the two points merge. At this stage another gapless sideband approach from the left.

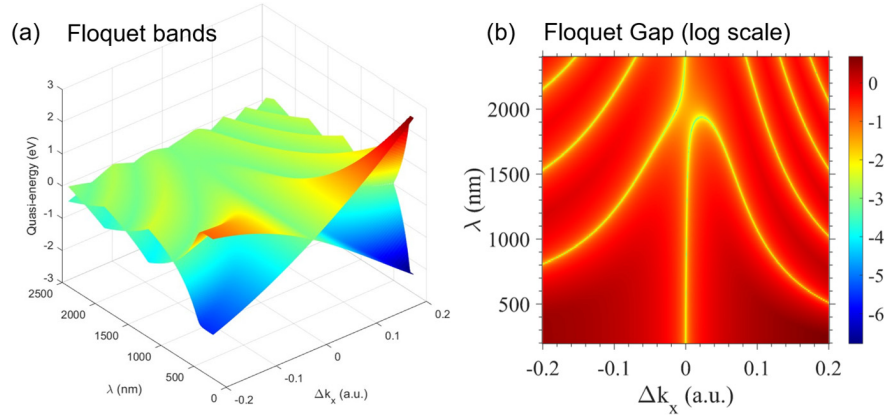

**FIG. S4.** Floquet band structure and Floquet quasi-energy gap along the $k_x$-axis for driving conditions similar to Fig. 1(a) in the main text along the $y$-axis, and *vs.* the driving wavelength. (a) Conduction and valence bands *vs.* driving wavelength, showing the motion of the Dirac point as the drive wavelength is tuned, up until it merges with a nearby replica band. (b) Same as (a), but showing the Floquet quasi-energy gap in logarithmic scale. Plot calculated with the same methodology as in Fig. 1 in the main text.

## 5. LONG WAVELENGTH ARPES

We numerically explore here ARPES signals from the Floquet pumped system with TDDFT, as in Fig. 3 in the main text, but in the case of longer wavelength driving, and weaker pulse peak power. Figure S5 shows ARPES spectra analogous with Fig. 3(a) in the main text, but for pumping wavelength of 4500nm, and peak power of $4\times10^{10}$ W/cm$^2$ (which is achievable with current technology[12]). The spectra clearly show that in the long wavelength pumping regime (since the vector potential is large), large movements of the Dirac point in graphene can still be observed even in lower peak powers.

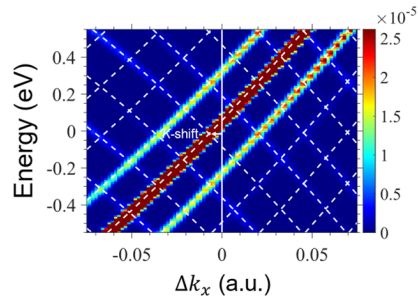

**FIG. S5.** *Ab-initio* TDDFT calculations of ARPES from light-driven graphene for linearly-polarized driving along $k_x$ at 4500nm and $4\times10^{10}$ W/cm$^2$. The spectrum is plotted along $k_x$ in the region of K, and is saturated for clarity. The overlayed dashed lines denote the Floquet quasi-energy bands obtained from the model in the same driving conditions. Arrows indicate shifting of the Dirac node and opening a gap at K.

## SUPPLEMENTARY REFERENCES